# Verification of the Crooks fluctuation theorem and recovery of RNA folding free energies

D. Collin[+], F. Ritort[#], C. Jarzynski[@], S. B. Smith[%], I. Tinoco[&], Jr. and C. Bustamante[%,*]

[+]*Merck & Co. Inc. , Automated Biotechnology Dpt., North Wales PA 19454*

[#]*Departament de Física Fonamental, Facultat de Física, Universitat de Barcelona, Diagonal 647, 08028 Barcelona, Spain*

[@]*T-13 Complex Systems, Los Alamos, National Lab, Los Alamos, NM 87545, USA*

[%]*Howard Hughes Medical Institute, University of California, Berkeley, CA 94720*

[&]*Department of Chemistry, University of California, Berkeley CA 94720, USA*

[*]*Departments of Physics and Molecular & Cell Biology, University of California, Berkeley CA 94720, USA*

*'First two authors contributed equally to this work'*

**The description of nonequilibrium processes in nano-sized objects, where the typical energies involved are a few times $k_B T$, is increasingly becoming central to disciplines as diverse as condensed-matter physics, materials science, and biophysics[1]. Major recent developments towards a unified treatment of arbitrarily large fluctuations in small systems are described by fluctuation theorems[2,3,4,5,6] that relate the probabilities of a system absorbing from or releasing to the bath a given amount of energy in a nonequilibrium process. Here we experimentally verify the Crooks Fluctuation Theorem[7] (CFT) under weak and strong nonequilibrium conditions by using optical tweezers[8] to measure the irreversible mechanical work during the unfolding and refolding of a small RNA hairpin[9] and an RNA three-helix junction[10]. We also show that the CFT provides a powerful way to obtain**

**folding free energies in biomolecules by determining the crossing between the unfolding and refolding irreversible work distributions. The method makes it possible to obtain folding free energies in nonequilibrium processes that dissipate up to $\sim 50 k_B T$ of the average total work exerted, thereby paving the way for reconstructing free energy landscapes[11] along reaction coordinates in nonequilibrium single-molecule experiments.**

Small systems thermodynamics involves the study of all energy exchange processes between a small system and its environment in the form of heat, work or mass. The CFT[7] predicts a symmetry relation in the work fluctuations when a system is driven away from thermal equilibrium by the action of an external perturbation. This theorem is based on the assumption that dynamics is microscopically reversible and therefore its experimental evaluation in small systems is crucial to better understand the foundations of nonequilibrium physics[12]. A consequence of the CFT is Jarzynski's equality[13] (JE), which relates the equilibrium free-energy difference $\Delta G$ between two equilibrium states to an exponential average (denoted by $<...>$) of the work done on the system, $W$, taken over an infinite number of repeated nonequilibrium experiments, $\exp(-\Delta G / k_B T) = < \exp(-W / k_B T) >$. In 2001, Hummer and Szabo[11] developed a formalism for reconstructing free energy profiles or potentials of mean force[14] along reaction coordinates, using nonequilibrium single-molecule pulling experiments, and they illustrated their method using data on the extraction of individual bacteriorhodopsin molecules. Recently, the JE was experimentally tested in single-molecule force experiments[15] on the P5ab RNA hairpin[16,17], a molecule that could be folded-unfolded quasi-reversibly. For processes that occur far from equilibrium, however the JE is hampered by large statistical uncertainties, arising from the sensitivity of the exponential average to rare events[18,19] (low values of $W$). Although for quasistatic processes the equality $<W> = \Delta G$ holds, it is often difficult in practice to extract unfolding free energies using small loading rates (below a few pN·s$^{-1}$) due to spatial drift in various components of the manipulation instrument. Drift effects decrease noticeably for larger pulling speeds, making it possible to obtain more reliable experimental data and better statistics (doing a large number of pulls), but at the expense of a more irreversible unfolding process. Here we show that significant improvements can be obtained by using the CFT, which provides a less sensitive and more rapidly converging method to extract equilibrium free energies from non-equilibrium processes.

Applied to the mechanical unfolding of a macromolecule, the CFT quantifies the amount of hysteresis observed in the values of the irreversible work obtained along unfolding and refolding paths. Let $P_U(W)$ denote the probability distribution of the values of the work performed on the molecule in an infinite number of pulling experiments along the unfolding (U) process, and define $P_R(W)$ analogously for the reverse (R) process. Let us now assume the following two conditions: 1) the unfolding and refolding processes are related by time-reversal symmetry, i.e. in our experiments, the optical trap used to manipulate the molecule is moved at the same speeds during the unfolding and refolding paths; 2) the molecule starts always in an equilibrium state (folded) in the unfolding process and in an equilibrium denatured extended state (unfolded) in the refolding process. The CFT[7] then predicts that

$$\frac{P_U(W)}{P_R(-W)} = \exp\left(\frac{W - \Delta G}{k_B T}\right) \qquad (1)$$

where $\Delta G$ is the free-energy change between final and initial states, equal to the reversible work associated with this process. Note that the CFT does not require that the system reaches the final equilibrium states immediately after the unfolding and refolding processes have been completed. It only requires that the control parameter attains its final value at the end of the process in either case, and that the system equilibrates to a well-defined state, consistent with the final value of the parameter. This latter process occurs without change of the control parameter, and therefore contributes no work. In principle, $W$ is an integral over the external variation of a control parameter[7], e.g. the position of the optical trap or the time (as pointed out by Hummer and Szabo[11]). However, in the present situation $W$ is well approximated by the familiar force-vs-extension integral,

$$W = \sum_{i=1}^{N_s} F_i \Delta x_i \qquad (2)$$

where $x_i$ is the distance between the ends of the molecule and $N_s$ is the number of intervals used in the sum (see[11] for a thorough discussion of this issue). Relation (1) quantifies hysteretic effects in the pulling experiment: work values larger than $\Delta G$ occur most often along the unfolding path while (absolute) values smaller than $\Delta G$ occur more often along the refolding path. The CFT, Eq. (1), states that although $P_U(W)$, $P_R(-W)$ depend on the pulling protocol, their ratio depends only on the value of $\Delta G$. Thus the value of $\Delta G$ can be determined once the distributions are known. In particular, the two distributions cross at $W = \Delta G$,

$$P_U(W) = P_R(-W) \Rightarrow W = \Delta G \qquad (3)$$

regardless of the pulling speed. Although the simple identity (3) already gives an estimate of $\Delta G$, it is not necessarily very precise as it only uses the local behavior of the distribution around $W = \Delta G$. Using the whole work distribution increases the precision of the free energy estimate[20]. In particular, as we show below, when the overlapping region of work values between the unfolding and refolding distributions is too narrow (as may happen for large values of the average dissipated work, defined as $\langle W_{dis} \rangle = \langle W \rangle - \Delta G$) the use of Bennett's acceptance ratio method[21] makes it possible to extract accurate estimates of $\Delta G$ using the CFT (see the Supplementary material). Below we first test the CFT for a case in which the unfolding occurs not too far from equilibrium (average dissipated work less than $6k_BT$). Next we extend this test to the case in which unfolding occurs very far from equilibrium (average dissipated work in the range $30-50 k_B T$). Finally we apply the CFT to obtain the difference in free energy between a wild type three-helix junction and a mutant molecule differing only by one base pair, and the free energy of stabilization of the wild type in the presence of $Mg^{2+}$ ions.

To experimentally test the validity of the CFT we used an siRNA hairpin that targets the mRNA of the CD4 receptor of the Human Immunodeficiency Virus[9] and that unfolds irreversibly but not too far from equilibrium at accessible experimental pulling speeds (dissipated work values less than $6 k_B T$). Under these conditions, the unfolding and refolding work distributions overlap over a sufficiently large range of work values to justify the use of the direct method to experimentally test Eq. (1). The work done on the molecules during pulling and relaxation, respectively, is given by the areas below the corresponding force-extension curves (Fig.1). Unfolding and refolding work distributions at three different pulling speeds are shown in Fig.2. Irreversibility increases with the pulling speed and unfolding-refolding work distributions become progressively more separated. Note, however, that the unfolding and the refolding distributions cross at a value of the work $\Delta G = 110.3 \pm 0.5 k_B T$ that does not depend on the pulling speed, as predicted by Eq. (3). Moreover the work distributions also satisfy the CFT, Eq. (1) (see the Supplementary Information). We also notice that work distributions are compatible with, and can be fitted to, Gaussian distributions (data not shown). After subtracting the contribution arising from the entropy loss due to the stretching of the molecular handles attached on both sides of the hairpin ($\Delta G^{handles} = 23.8\ k_B T$) and of the extended single stranded RNA ($\Delta G^{ssRNA} = 23.7 \pm 1\ k_B T$) from the total work, $\Delta G^{(exp)} = 110.3 \pm 0.5 k_B T$, we obtain for the free energy of unfolding at zero force $\Delta G_0^{(exp)} = 62.8 \pm 1.5 k_B T = 37.2 \pm 1 kcal/mol$ (at 25˚C, in 100 mM Tris-HCl, pH 8.1, 1 mM EDTA), in excellent agreement with the result obtained using the Visual OMP from DNA software[22] $\Delta G_0^{(mfold)} = 38 kcal/mol$ (at 25°C, in 100 mM NaCl).

To extend the experimental test of the validity of the CFT to the very-far-from-equilibrium regime where the work distributions are no longer Gaussian, we then apply the CFT to determine: 1) the difference in folding free energy between an RNA molecule and a mutant that differ only by one base-pair, and 2) the thermodynamic stabilizing effect of $Mg^{2+}$ ions on the RNA structure. The RNA we consider is a three-helix junction of the 16S ribosomal RNA of *Escherichia coli*[10] that binds the S15 protein. The secondary structure of this RNA is a common feature in RNA structures[23,24,25] that plays, in this case, a crucial role in the folding of the central domain of the 30S ribosomal subunit. For comparison, and to verify the accuracy of the method, we have pulled the wild-type and a C•G to G•C mutant (C754G-G587C) of the three-helix junction. Fig. 3 depicts the unfolding and refolding work distributions for the wild type and mutant molecules (work values were binned into about 10-20 equally spaced intervals). For both molecules, the distributions display a very narrow overlapping region. In contrast with the hairpin distribution, the average dissipated work for the unfolding pathway is now much larger—in the range $20-40 k_B T$ — and the unfolding work distribution shows a large tail and strong deviations from Gaussian behaviour. Thus, these molecules are ideal to test the validity of Eq. (1) in the far from equilibrium regime. As shown in the inset of Fig. 3, the plot of the log ratio of the unfolding to the refolding probabilities vs. total work done on the molecule can be fitted to a straight line with a slope of 1.06, thus establishing the validity of the CFT (see Eq. 1) under far from equilibrium conditions. Our measurements reveal the presence of long tails in the work distribution $P_U(W)$ along the unfolding path and narrow work distributions $P_R(W)$ along the refolding path. These distributions complement each other, one being large where the other is small, thereby providing thermodynamically important information about the free energy landscape.

Bennet's acceptance ratio method gives $\Delta G^{(exp)} = 154.1 \pm 0.4\, k_B T$ and $\Delta G^{(exp)} = 157.9 \pm 0.2\, k_B T$ for unfolding the wild and mutant types respectively giving a difference between the two forms $\Delta\Delta G_0^{(exp)} = \Delta\Delta G^{(exp)} = 3.8 \pm 0.6\, k_B T$. After subtracting the (identical for both molecules) handle and RNA entropy loss contributions ($97 \pm 1\, k_B T$) we get $\Delta G_0^{(exp)} = 57 \pm 1.5\, k_B T$ (wild type) and $\Delta G_0^{(exp)} = 60.8 \pm 1.5\, k_B T$ (mutant), the error increasing due to the uncertainty in the contributions coming from the stretching of ssRNA. Free-energy prediction programs such as Mfold [26] and Visual OMP [22] give a $\Delta\Delta G_0^{(mfold)} = 2 \pm 2\, k_B T$ between the forms (at 25˚C and 100 mM NaCl). Thus, when combined with acceptance ratio methods, the CFT furnishes a method precise enough to determine the difference in the folding free energies of RNA molecules differing only by one base pair in 34 base pairs.

Finally we apply Eq.(1) to obtain the free energy of stabilization by $Mg^{2+}$ of the S15 three-helix junction. These values are often difficult to access by bulk methods because melting temperatures of tertiary folded RNAs are frequently higher than the boiling point of water, and $Mg^{2+}$ catalyzes the hydrolysis of RNA at elevated temperatures[27]. Fig. 4 depicts the work histograms in the presence and absence of $Mg^{2+}$ (at constant ionic strength); stretching contributions differ in the presence and absence of magnesium ions ($116.8\, k_B T$ and $97\, k_B T$, respectively). These values have been subtracted from the work data to properly compare the unfolding free energies of both molecules. The strong increase of irreversibility due to $Mg^{2+}$ can be seen in the large value of the average dissipated work (about $50\, k_B T$ along the unfolding reaction and $16\, k_B T$ along the refolding path). Applying Bennett's acceptance ratio method for the molecule in the presence of magnesium yields $\Delta G^{(exp)} = 205.5 \pm 1.5\, k_B T$ and (after subtracting the stretching contributions) gives $\Delta G_0^{(exp)} = 88.7 \pm 2.5\, k_B T$ for the unfolding reaction of the wild-type junction in 4 mM MgCl$_2$. The difference in free energies of unfolding in the presence and absence of $Mg^{2+}$, $\Delta\Delta G_0^{(exp)} = -31.7 \pm 2\, k_B T$, gives the free 8energy of stabilization associated with the binding of $Mg^{2+}$ ions to the S15 three-helix junction both through specific and non-specific (shielding) interactions.

How reliable is the current method to recover free energies in other systems (e.g. proteins) and/or using other techniques (e.g. AFM)? Our method works for soft traps but is probably limited to molecules that dissipate less than $100 k_B T$. Extending this approach to much stiffer AFM cantilevers pulling proteins[28] is a question of current enquiry in our laboratory. Finally, our method might have limitations when applied to molecular interactions such as ligand binding or macromolecular assembly processes, if it is not possible to identify a molecular extension change that corresponds to the reaction coordinate of the system.

Fluctuation theorems establish fundamental relationships about the energy exchanged by a system and its environment in a non-equilibrium process. Testing their validity, and exploring their limitations, adds to our understanding of the thermodynamics of small systems. Ultimately, these theorems may clarify the role of fluctuations in the operation of biological motors, and have implications for the design of nanodevices operating far from thermal equilibrium. Finally, our results show that single molecule methods, largely developed for biochemical studies, can play a significant role in unraveling fundamental physical relationships and that the latter, in turn, can be used to better characterize the thermodynamics of non-equilibrium biochemical processes.

**Methods.**

**Sample preparation**. The RNA molecules were prepared as previously described by Liphardt et al.[15]. Each DNA sequence corresponding to the three different RNAs was cloned separately into pBR322 vector between *Eco*R I and *Hin*d III sites. A polymerase chain reaction (PCR) was used to amplify a DNA sequence containing an upstream T7 promoter, the RNA sequence of interest and flanking DNA sequences corresponding to the "handles". The handles correspond to a sequence of pBR322 (NCBI ID "J01749") from nucleotide 3838 to 1 and from 29 to 629, respectively. The three RNA sequences were transcribed *in vitro* using T7 RNA polymerase[29]. Two DNA handles were synthesized by PCR. The DNA handle upstream of the RNA was biotinylated at the 3'-end, whereas a digoxigenin moiety was attached to the 5'-end of the other handle. The RNA and two DNA handles were annealed by heating samples to 85 $^o$C, followed by a slow cooling down to room temperature. The RNA hairpin was pulled in 100mM Tris-HCl, pH 8.1, 1 mM EDTA buffer. The S15 three-helix junction and the mutant have been pulled in 62 mM KCl, 10 mM HEPES pH 7.8 buffer. In 4 mM MgCl$_2$ the KCl concentration was adjusted to 50 mM to work at the same ionic strength than in the absence of Mg$^{2+}$.

**Work measurements.** Work probability distributions were obtained from many force extension curves for a given molecule and aligned to a worm-like chain curve that best fitted the force-extension data at forces below the range of forces where the molecule unfolds. This procedure minimizes the effect of machine drift on the measured work values. For the worm-like chain fits we used $P \approx 10 nm$ and $P \approx 1 nm$ for the persistence lengths of the DNA/RNA hybrid handles and ssRNA respectively. Work values where integrated along the following range of extension: [355 nm, 380 nm] for the hairpin and [326 nm, 392 nm] for the S15 three-helix junction without magnesium (wild and mutant) and [337 nm, 398 nm] with magnesium. The free-energy contributions from stretching the handles and the ssRNA were then obtained by numerical integration of the worm-like chain reference curves using the values for the persistence and contour lengths of the polymers. To estimate the free energy of unfolding at zero, $\Delta G_0^{(\exp)}$, we subtract the free-energy contribution of the hybrid handles and ssRNA from the total reversible work across the transition, $\Delta G^{(exp)}$, by using the expression[16,30] $\Delta G_0^{(exp)} = \Delta G^{(exp)} - \Delta G^{handles} - \Delta G^{ssRNA}$, where $\Delta G^{handles}, \Delta G^{ssRNA}$ are the entropy loss contributions due to the stretching of the molecular handles attached on both sides of the hairpin and of the extended single stranded RNA respectively.

**Table 1**

| Molecule | $W_m^U$ | $W_m^R$ | $\sigma_U$ | $\sigma_R$ | $W_{cum}^{(2)}$ | $W_J^U$ | $W_J^R$ | $W_J^{(est)}$ | $\Delta G^{(exp)}$ | $\Delta G_0^{(exp)}$ | $W_{dis}^U$ | $W_{dis}^R$ | $R_U$ | $R_R$ |
|---|---|---|---|---|---|---|---|---|---|---|---|---|---|---|
| (*)Hairpin (1.5pN/s) | 110.9 | 108.7 | 2.35 | 2.21 | 109.7 (0.2) | 107.4 (0.7) | 110.9 (0.2) | 109.1 (0.5) | **110.0 (0.2)** | *62.5 (1.2)* | 0.9 | 1.3 | 3.1 | 1.9 |
| Hairpin (7.5pN/s) | 113.8 | 106.6 | 2.63 | 2.84 | 110.3 (0.2) | 109.7 (0.7) | 110.9 (0.5) | 110.3 (0.5) | **110.3 (0.5)** | *62.8 (1.5)* | 3.5 | 3.7 | 0.98 | 1.10 |
| Hairpin (20pN/s) | 115.7 | 104.1 | 3.2 | 3.5 | 110.1 (0.2) | 110.2 (0.7) | 108.6 (0.2) | 109.4 (0.4) | **110.2 (0.6)** | *62.9 (1.6)* | 5.4 | 6.2 | 0.94 | 0.98 |
| S15(wild,no Mg) | 191.3 | 145.9 | 11.3 | 2.9 | 158.7 (0.8) | 155.2 (1.4) | 149.3 (0.2) | 152.2 (0.7) | **154.1 (0.4)** | *57.0 (1.5)* | 36.3 | 9.1 | 1.75 | 0.46 |
| S15(mut,no Mg) | 176.5 | 153.4 | 10.6 | 2.1 | 156.0 (0.4) | 152.4 (5.0) | 155.7 (0.2) | 154.1 (0.3) | **157.9 (0.2)** | *60.8 (1.5)* | 18.6 | 4.5 | 3.02 | 0.49 |
| S15(wild,Mg) | 256.4 | 190.3 | 12.2 | 5.0 | 213.0 (1.3) | 207.0 (4.0) | 199.8 (0.6) | 203.6 (2.0) | **205.5 (1.5)** | *88.7 (2.5)* | 50.9 | 15.2 | 1.46 | 0.82 |

Table 1. Summary of results obtained for all molecules. $W_m^{U,R}$, $\sigma_{U,R}$, $W_J^{U,R}$ are the average total work, standard deviation and predictions obtained by using the JE along the unfolding (U) and refolding (R) paths. $W_{cum}^{(2)}$ is the estimate obtained by Hummer[31], $W_{cum}^{(2)} = (W_m^U + W_m^R)/2 - (\sigma_U^2 - \sigma_R^2)/12k_BT$, which gives the leading correction to the linear response prediction, $W_J^{(est)}$ is the average of the estimates obtained by using the JE along the unfolding and refolding paths $W_J^{(est)} = (W_J^U + W_J^R)/2$, $\Delta G^{(exp)}$ is our best estimate obtained by using the acceptance ratio method (in bold font type), $\Delta G_0^{(exp)}$ is the final estimate for the unfolding free energy at zero force after subtracting the handles contribution (in italicized bold font type), $W_{dis}^{U,R} = |W^{U,R} - \Delta G^{(exp)}|$ is the average dissipated work (for the analysis of the hairpin data we took $\Delta G^{(exp)} = 110.3$ for all pulling rates) and $R_{U,R} = \frac{\sigma_{U,R}^2}{2k_BTW_{dis}^{U,R}}$ is a parameter that is equal to 1 for Gaussian work distributions[17]. Statistical errors are given for the JE and the crossing estimates. These were obtained using the bootstrap method. All work values (except $\Delta G_0^{(exp)}$) include the handle and RNA stretching contributions and are given in units of $k_BT$ at $T = 298K$. In parentheses we indicate the errors in $k_BT$ units. (*) Data for the RNA hairpin at 1.5pN/s are also included for completion, however at such low loading rates drift effects are very large and data is very noisy as revealed by the values of $R_U$ and $R_R$ which differ too much from 1.

# Figure captions

Figure 1: Force-extension curves. The stochasticity of the unfolding and refolding process is characterized by a distribution of unfolding or refolding work trajectories. Five unfolding (orange) and refolding (blue) force-extension curves for the RNA hairpin are shown (loading rate of 7.5 pN/s). The blue area under the curve represents the work returned to the machine as the molecule switches from the unfolded to the folded state. The RNA sequence is shown as an inset.

Figure 2: Test of the CFT using an RNA hairpin. Work distributions for RNA unfolding (continuous lines) and refolding (dashed lines). We plot negative work, $P_R(-W)$, for refolding. Statistics: 130 pulls and 3 molecules (r = 1.5 pN/s), 380 pulls and 4 molecules (r = 7.5 pN/s), 700 pulls and 3 molecules (r = 20.0 pN/s), for a total of 10 separate experiments. Good reproducibility was obtained among molecules (see Figure S2 in Supplementary Information). Work values were binned into about 10 equally spaced intervals. Unfolding and refolding distributions at different speeds show a common crossing around $\Delta G = 110.3 \ k_B T$.

Figure 3: Free-energy recovery and test of the CFT for non-Gaussian work distributions. Experiments were carried out on the wild type and mutant S15 three-helix junction without $Mg^{2+}$. Unfolding (continuous lines) and refolding (dashed lines) work distributions. Statistics: 900 pulls, 2 molecules (wild type, indigo color); 1200 pulls, 5 molecules (mutant type, orange color). Crossings between distributions are indicated by black circles. Work histograms were found to be reproducible among different molecules (error bars indicating the range of variability). (Inset) Test of the CFT for the mutant. Data have been linearly interpolated between contiguous bins of the unfolding and refolding work distributions.

Figure 4: Use of CFT to extract the stabilizing contribution of $Mg^{2+}$ to the free energy of the S15 three-helix junction (wild type). Unfolding (continuous lines) and refolding (dashed lines) work distributions. Green curves: 450 pulls, 2 molecules in $Mg^{2+}$; Indigo color: 900 pulls, 2 molecules without $Mg^{2+}$. Crossings between distributions are indicated by black circles. Work histograms are reproducible between the molecules (error bars indicating the range of variability). (Inset) Histograms in logarithmic scale showing (thick black bars) the regions of work values where unfolding and refolding distributions are expected to cross each other by Bennet's acceptance ratio method (Supplementary Information).

1. Liphardt, J., Bustamante, C. & Ritort, F. The nonequilibrium thermodynamics of small systems. *Physics Today.* **58,** 43-48 (2005).

2. Evans, D. J., Cohen, E. G. & Morriss, G. P. Probability of second law violations in shearing steady states. *Phys. Rev. Lett.* **71,** 2401-2404 (1993).

3. Gallavotti, G. & Cohen, E. G. D. Dynamical ensembles in nonequilibrium statistical mechanics. *Phys. Rev. Lett.* **74,** 2694-2697 (1995).

4. Ciliberto, S. & Laroche, C. An experimental test of the Gallavotti-Cohen fluctuation theorem. *J. Phys. IV (France)* **8**, Pr6-215 (1998).

5. Evans, D. J. & Searles, D. J. The Fluctuation theorem. *Adv. Phys.* **51,** 1529-1585 (2002).

6. Wang, G. M., Sevick, E. M., Mittag, E., Searles D. J. & Evans, D. J. Experimental demonstration of violations of the second law of thermodynamics for small systems and short timescales. *Phys. Rev. Lett.* **89,** 050601 (2002).

7. Crooks, G. E. Entropy production fluctuation theorem and the nonequilibrium work relation for free-energy differences. *Phys. Rev. E.* **60,** 2721-2726 (1999).

8. Smith, S. B., Cui, Y. & Bustamante, C. An optical-trap force transducer that operates by direct measurement of light momentum. *Methods. Enzymol.* **361**, 134 (2003).

9. McManus, M. T., Petersen, C. P., Haines, B. B., Chen, J. & Sharp, A. P. Gene silencing using micro-RNA designed hairpins. *RNA* **8**, 842-850 (2002).

10. Serganov. A. et al. Role of conserved nucleotides in building the 16S rRNA binding site for ribosomal protein S15. *J. Mol. Biol.* **305**, 785-803 (2002).

11 . Hummer, G. & Szabo, A. Free-energy reconstruction from nonequilibrium single molecule experiments. *Proc. Natl. Acad. Sci. USA* **98,** 3658-3661 (2001).

12. Ritort, F.Work fluctuations, transient violations of the second law and free-energy recovery methods. *Seminaire Poincare* **2,** 193–226 (2003).

13. Jarzynski, C. Nonequilibrium equality for free energy differences. *Phys. Rev. Lett.* **78**, 2690-2693 (1997).

14. Park, S. & Schulten, K. Calculating potentials of mean force from steered molecular dynamics simulations. *J. Chem. Phys.* **120,** 5946-5961 (2004).

15. Liphardt, J., Onoa, B., Smith S. B., Tinoco, I. Jr. & Bustamante, C. Reversible unfolding of single RNA molecules by mechanical force. *Science* **292,** 733-737 (2001).

16. Liphardt, J., Dumont, S., Smith S. B., Tinoco, I. Jr. & Bustamante, C. Equilibrium information from nonequilibrium measurements in an experimental test of the Jarzynski equality. *Science* **296,** 1832-1835 (2002).

17. Ritort, F., Bustamante, C. & Tinoco, I. Jr. A two-state kinetic model for the unfolding of single molecules by mechanical force. *Proc. Nat. Acad. Sci. USA* **99,** 13544-13548 (2002).

18. Zuckerman, D. M. & Woolf, T. B. Theory of systematic computational error in free energy differences. *Phys. Rev. Lett.* **89,** 180602 (2002).

19. Gore, J., Ritort, F. & Bustamante, C. Bias and error in estimates of equilibrium free-energy differences from nonequilibrium measurements. *Proc. Nat. Acad. Sci. USA* **100,** 12564-12569 (2003).

20. Shirts, R., Bair, E., Hooker, G. & Pande, V. S. Equilibrium free energies from nonequilibrium measurements using maximum likelihood methods. *Phys. Rev. Lett*. **91**, 140601 (2003).

21. Bennett, C. H. Efficient estimates of free energy differences from Monte Carlo data. *J. Comp. Phys*. **22**, 245-268 (1976)

22. SantaLucia J, Jr. & Hicks, D. The thermodynamics of DNA structural motifs. *Ann. Rev. Biophys. Biomol. Struct.* **33,** 415-440 (2004).

23. Robertus, D. J. et al. Structure of yeast phenylalanine tRNA at 3A resolution. *Nature* 250, 546-551 (1974)

24. Long, D. M. & Uhlenbeck, O. C. Self-cleaving catalytic RNA. *FASEB J*. **7,** 25-30 (1993).

25. Cate, J. H. & Doudna, J. A. Metal binding sites in the major groove of a large ribozyme domain. *Structure* **4,** 1221-1229 (1996).

26. Zuker, M. Mfold web server for nucleic acid folding and hybridization predictions. *Nucleic Acids Res*. **31**, 3406-3415 (2003).

27. Turner, D. H. in *Nucleic Acids: Structures, Properties and Functions*, Bloomfield, D. A., Crothers, D. M. & Tinoco, I. Jr., Eds. (University Press, New York, 2000), Chapter 7.

28. Carrion-Vazquez, M., Oberhauser, A. F., Diez, H., Hervas, R., Oroz, J., Fernandez, J. & Martinez-Martin, D. (in press). Protein nanomechanics studied by AFM single-molecule force spectroscopy. *In: Arrondo, J. L. R., and Alonso, A (eds.) "Emerging Techniques in Biophysics" (Biophysics Monograph Series). Springer-Verlag, Heidelberg.*

29. Milligan, J. F., Groebe, D. R., Witherell, G. W. & Uhlenbeck, O. C. Oligoribonucleotide synthesis using T7 RNA polymerase and synthetic DNA templates. *Nucleic Acids Res.* **15**, 8783-8798 (1987).

30. Manosas, M. & Ritort, F. Thermodynamic and kinetic aspects of RNA pulling experiments. *Biophys. J.* **88,** 3224-3242 (2005).

31. Hummer, G. Fast-growth thermodynamics integration: Error and efficiency analysis. *J. Chem. Phys.* **114,** 7330-7337 (2001).

**Supplementary Information** accompanies the paper on *Nature*'s website (http://www.nature.com).

**Acknowledgements.** We are grateful to G. Hummer and A. Szabo for many useful discussions and to G. E. Crooks, D. Chandler and J. Liphardt for a careful reading of the manuscript. F. R. has been supported by the Spanish Research council and the Catalan Government (Distinció de la Generalitat). C. J. has been supported by an NIH grant and the U.S. Department of Energy. I. T. has been supported by an NIH grant. C. B . has been supported by the Howard Hugues Medical Institute and the David and Lucile Packard Foundation.

**Competing interests statement.** The authors declare that they have no competing financial interests.

**Correspondence** and requests for material should be addressed either to C. B or F. R (E-mail: carlos@alice.berkeley.edu, ritort@ffn.ub.es ).

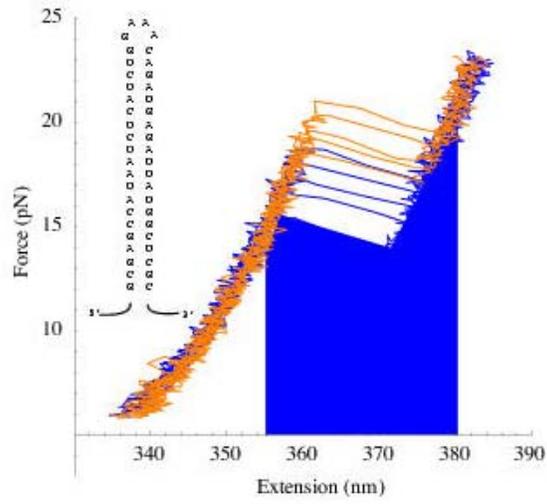

Figure 1

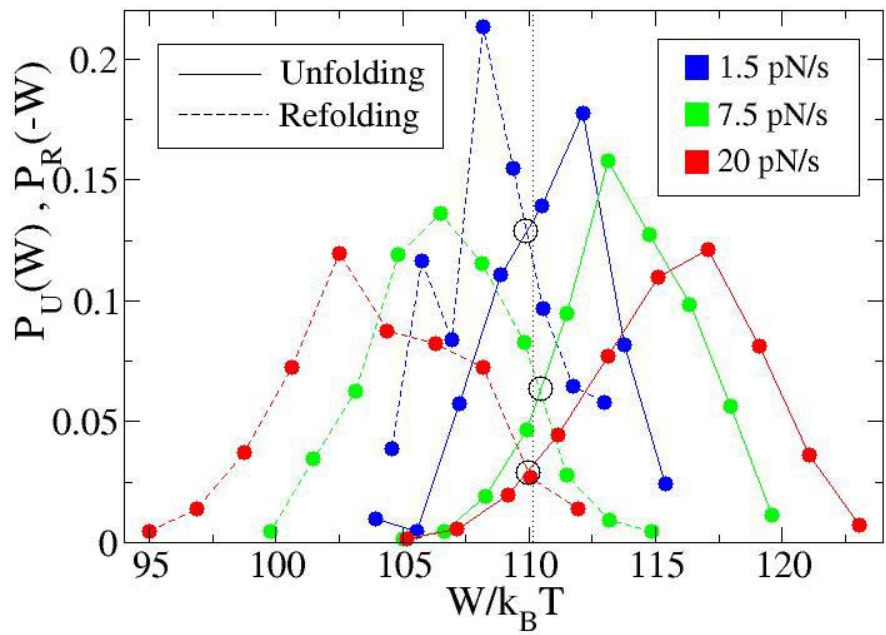

Figure 2

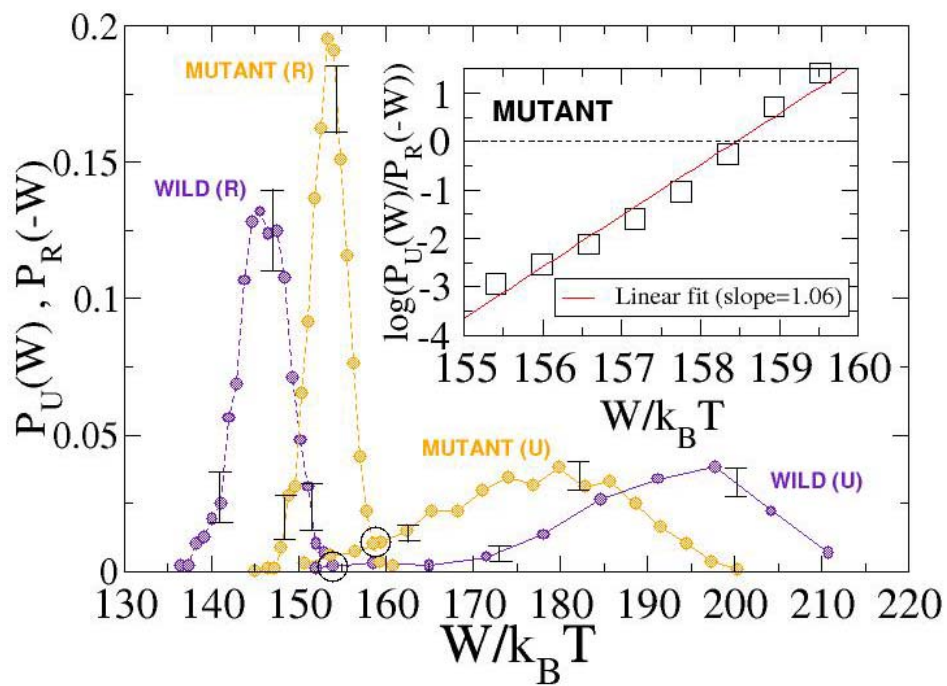

Figure 3

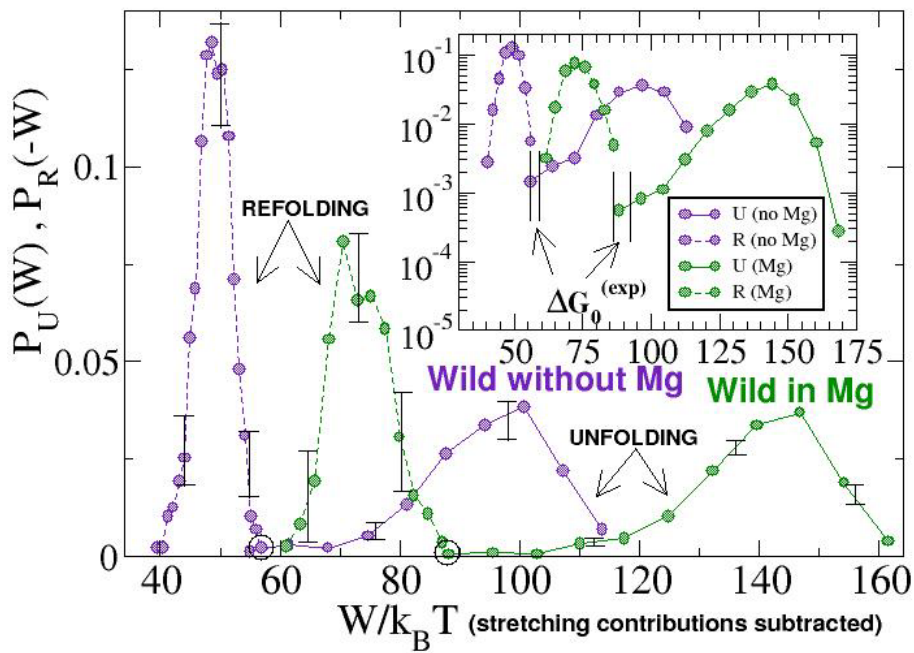

Figure 4